\documentclass[twocolumn,showpacs,preprintnumbers,amsmath,amssymb]{revtex4}

\usepackage{graphicx}
\usepackage{dcolumn}
\usepackage{bm}

\newcommand{\Buzek}{Bu\v{z}ek}

\newcommand{\real}{\mathbf{R}}
\newcommand{\complex}{\mathbf{C}}

\newcommand{\Hilbert}{\mathcal{H} }
\newcommand{\state}{\mathcal{S}(\mathcal{H})}

\newcommand{\Tr}{\mathrm{Tr}}
\newcommand{\Exp}{\mathrm{E}}

\newcommand{\im}{\mathrm{Im}}
\newcommand{\rmd}{\mathrm{d}}
\newcommand{\dtheta}{\mathrm{d}\theta}
\newcommand{\dx}{\mathrm{d}x}
\newcommand{\half}{\frac{1}{2} }

\begin{document}

\title{
Bayesian prediction of the Gaussian states from $n$ sample
}

\preprint{APS/123-QED}

\title{Bayesian prediction from $n$ sample\\ on the Gaussian state model}

\author{Fuyuhiko Tanaka}
\email{ftanaka@stat.t.u-tokyo.ac.jp}
\author{Fumiyasu Komaki}%
\affiliation{%
Department of Mathematical Informatics, University of Tokyo, 7-3-1 Hongo, Bunkyo-ku, Tokyo, 113-8656  Japan
}%

\date{\today}

\begin{abstract}
%
%
Recently quantum prediction problem was proposed in the Bayesian framework~\cite{FT3}.
It is shown that Bayesian predictive density operators  are the best predictive density operators 
when we evaluate them by using the average relative entropy based on a prior.
As an illustrative example, we treat the Gaussian states family adopting the Gaussian distribution as a prior
 and give the Bayesian predictive density operator with the heterodyne measurement fixed.
We show that it is better than the plug-in predictive density operator
 based on the maximum likelihood estimate by calculating each average relative entropy.
\end{abstract}

\pacs{03.67.-a,03.65.Yz}
\keywords{Gaussian states, quantum prediction, relative entropy}
\maketitle


\section{\label{sec:level1}Introduction}
 In quantum statistics, problems of statistical inference and state estimation has received a lot of attention
 over the past several years with recent developments of experimental techniques.
Historically speaking, parameter estimation problem on quantum systems dates back to
 a quarter century, when Helstrom, Holevo, and other researchers vigorously investigated the topic and gave some extension  of mathematical statistical concepts 
on classical probability.

Bayesian approach for quantum statistics has also been investigated~\cite{Helstrom, Holevo}.
Jones~\cite{Jones1991} has derived a quantum Bayes rule for pure states with the uniform prior.
Later, \Buzek \ \textit{et al.}~\cite{Buzek1998} pointed out that 
it can be applied to mixed states with purification ansatz.
Schack \textit{et al.}~\cite{Schack2001} extended his result to a more general framework of exchangeable states.
They showed that a quantum state after a measurement can be interpreted as the state averaged over the posterior.
\Buzek \ \textit{et al.}~\cite{Buzek1998} recommended to use Bayesian technique especially when the sample size of
 experimental data is small.
They proposed to use a posterior state corresponding to a posterior distribution in classical counterparts. 



From the viewpoints of information quantity and Bayes rule, however, Bayesian estimation on quantum states has not been fully discussed.
Performances of the Bayesian approach compared with other approach such as the maximal likelihood method have not been discussed theoretically.
Tanaka and Komaki showed that  the Bayesian method has better performance than the plug-in method when exchangeable states are considered~\cite{FT3}.
In the present paper, we review it and calculate the Bayesian predictive density operator for the Gaussian states family with the heterodyne measurement.


\section{Preliminary}

We briefly summarize some notations of quantum measurement. 
Let $\Hilbert $ be a separable (possibly infinite dimensional) Hilbert space of a quantum system.
An Hermitian operator $\rho $ on $\Hilbert $ is called a \textit{state} or \textit{density operator} if it satisfies,
\[
\Tr \rho = 1, \quad \rho \geq 0.
\]
We denote the set of all states on $\Hilbert $ as $\state $. 

Let $\Omega $ be a space of all possible outcomes of an experiment (e.g., $\Omega = \real^{n}$)
 and suppose that a $\sigma$-algebra $\mathcal{B}  := \mathcal{B}(\Omega) $ of subsets of $\Omega $ is given. 
An affine map $\mu $ from $\state $ into a set of probability distributions on $\Omega$, $\mathcal{P}$=
 $\{ \mu (\dx )  \}$ is called a \textit{measurement}.
There is a one-to-one correspondence between a measurement
 and a resolution of the identity~\cite{Holevo}.
A map from $\mathcal{B}$ into the set of positive Hermitian operators 
\[ 
 M: B \mapsto M(B),
\]
where $M $ satisfies
\begin{gather}
 M(\phi) = O, M(\Omega )=I, \nonumber \\ 
 M(\cup_{i} B_i ) = \sum_i M(B_i),  \quad B_i \cap B_j = \phi,  \quad  \forall B_i \in \mathcal{B }, \nonumber  
\end{gather}
is called a \textit{positive operator valued measure} (\textit{POVM}).
Any physical measurement can be represented by a POVM.


Now we describe our setting of state estimation.
Assume that a state $\rho_{\theta} $ on $\Hilbert $ is characterized by an unknown finite-dimensional parameter $\theta \in \Theta \subset  \real^p$.

A quantum state for $n$ systems, $\rho^{(n)}$, is described on the $n$-fold tensor product Hilbert space $\Hilbert ^{\otimes n}$.
Suppose that a system composed of $n$+$m$ subsystems is given and that a measurement is performed only for selected $n$ subsystems with the other $m$ subsystems
 left.
Then, the measurement is described by $\{ M_x \!\otimes \!I \} $, where $\{ M_x \}$ is a POVM on $\Hilbert^{\otimes n}$ and 
$I$ is the identity operator on $\Hilbert^{\otimes m} $.

Our aim is to estimate the true state $\sigma_{\theta }:= \rho_{\theta }^{\otimes m}$ of the remaining $m$ subsystems
 by using a measurement $\{ M_x \}$ on the selected $n$ subsystems
 $\rho_{\theta}^{\otimes n}$. 
We fix an arbitrarily chosen measurement.
Note that this measurement is not necessarily in the form of a tensor product $M_x^{\otimes n}$, which represents a repetition of the same measurement $M_x$ for each system.
Thus, all possible measurements on $n$ subsystems, which may use entanglement,  are considered.

The performance of a predictive density operator $\hat{\sigma }(x) $ is evaluated by the relative entropy 
$D(\sigma_{\theta} || \hat{\sigma} (x)) $, a quantum analogue of the Kullback-Leibler divergence in classical statistics.
The quantum relative entropy from $\rho $ to $\sigma$ is defined by
\begin{equation}
 D(\rho || \sigma ):= \Tr [ \rho (\log \rho - \log \sigma ) ]. \label{eq:rel}
\end{equation}
It satisfies the positivity condition $D(\rho || \sigma ) \geq 0$ and $D(\rho || \sigma) =0 \Leftrightarrow \rho = \sigma$.
Thus, it can be used as a measure for the goodness of state estimation.

There are mainly two approaches on inference of state $\sigma_{\theta }$ for the parametric model above.
One approach is to use  $\sigma_{\hat{\theta}(x) } $, where $\hat{\theta}(x)$ is an estimator of $\theta $,
 depending on the observation $x$.
The other approach corresponds to the Bayesian predictive density approach in classical statistics~\cite{Jones1991, Buzek1998}.
We shall briefly review the idea.
First, we assume a probability density $\pi (\theta )$ on the parameter space.
In mathematical statistics $\pi (\theta)$ is usually called a \textit{prior density}.
When there is no knowledge about parameter $\theta $, which is often called \textit{noninformative},
  several people have discussed what kind of prior should be used \cite{Slater1997}, \cite{Braunstein1994}.
From the data $x$ obtained from a measurement $\{ M_x \}$, a posterior distribution $\pi(\theta | x)$ is constructed
 as
\[
 \pi(\theta | x) := \frac{p^{M}(x | \theta)\pi(\theta ) }{\int \dtheta  \ p^{M}(x | \theta)\pi(\theta ) },
\]
where $p^{M}(x| \theta ) = \Tr \rho^{\otimes n}_{\theta }M_x $.
Next, taking an average of $\sigma_{\theta }$ with $\pi (\theta | x)$, one can obtain the Bayesian estimator 
\[
  \sigma_{\pi}(x) = \int \dtheta \ \sigma_{\theta} \pi(\theta | x). 
\]
We call this state estimator, as in classical statistics, a \textit{Bayesian predictive density operator}.
In order to distinguish two estimators we call $\sigma_{\hat{\theta}}$, an estimator based on $\hat{\theta }$, 
a \textit{plug-in predictive density operator}.

If we assume a prior probability density $\pi(\theta ) $ on the parameter space $\Theta $, the mixture state is given by
\begin{equation}
 \rho^{(n)} := \int \dtheta \ \pi(\theta) \ \rho_{\theta}^{\otimes n}  \label{eq:exca}.
\end{equation}
A state of the form (\ref{eq:exca}) is called an \textit{exchangeable state}~\cite{Schack2001}, and arises, e.g., if each subsystem is  
prepared in the same unknown way, as in quantum state tomography.
In a quantum exchangeable model~(\ref{eq:exca}), as Schack \textit{et al.}~\cite{Schack2001} showed, 
a posterior distribution $\pi (\theta | x)$ naturally arises.

Tanaka and Komaki show that Bayesian predictive density operators are better than plug-in predictive density operators~\cite{FT3}.

\noindent{\it Theorem. 1}\\
Suppose that we perform a measurement for selected $n$ subsystems $\rho_{\theta }^{\otimes n}$of 
 a system $\rho_{\theta }^{\otimes (n+m) }$ composed of $n+m$ subsystems in order to estimate the remaining $m$ subsystems $\sigma_{\theta } = \rho_{\theta }^{\otimes m}$. 
The true parameter value $\theta $ is unknown and a prior probability density $\pi (\theta)$ is assumed.
Let $\hat{\sigma}(x) $ be any predictive density operator,
where $x$ is an outcome of a measurement $\{ M_x \}$ for the $n$ subsystems.
Performance of a predictive density operator $\hat{\sigma }(x)$ is measured with the average relative entropy
\[
 \Exp^{\pi }\Exp^{M} [D(\sigma_{\theta } || \hat{\sigma }(x) ) ] \!\!= \!\! \int\! \dtheta \pi(\theta ) \!\!\! \int\!\! \dx p^{M}\!(x| \theta)
 D(\sigma_{\theta } || \hat{\sigma }(x) ) 
\] 
 from the true state $\sigma_{\theta }$.
 Then, the Bayesian predictive density operator $\sigma_{\pi}(x)$
based on the observation $x$ and the prior $\pi (\theta )$ is the best predictive density operator.\\

\noindent\textit{Remark.}\\
In classical statistics, Aitchison~\cite{Ait1975} showed that
the Bayesian predictive density $p_{\pi}(y|x)$ has better performance under the Kullback-Leibler divergence
 than any plug-in predictive density $p(y| \hat{\theta })$ when a proper prior $\pi (\theta) $ is given.
Theorem $1$ is the corresponding result for quantum predictive density operators.

In different setting, Krattenthaler and Slater obtained a similar result as a quantum version of the Aitchison's result~\cite{Kratten1996}. 
While they consider a prior density $\pi(\theta)$ with respect to an unknown state,
 we consider a posterior density $\pi(\theta |x)$ with respect to a post-measurement state.


%
%

\section{Prediction of unknown Gaussian state from one sample}

We consider the prediction problem of  the Gaussian states family below (See, e.g., Holevo~\cite{Holevo} for the Gaussian states family).
\begin{eqnarray}
 \mathcal{M} &:=& \left\{ \rho_{\theta, N}  :\  \theta \in \complex \right\}, \label{eq:model}\\
 \rho_{\theta, N} &:=& \frac{1}{\pi N}  \int_{\complex} \exp \left( - \frac{ | \alpha -\theta |^2   }{N} \right) | \alpha \rangle \langle \alpha | 
  \rmd ^2 \alpha  \ \nonumber
\end{eqnarray}
and assuming that the photon expectation parameter $N (>0) $ is known. We omit $N$ unless otherwise necessary. 

The parameter estimation problem of the model (\ref{eq:model}) was investigated by Yuen and Lax~\cite{Yuen1973} and Holevo~\cite{Holevo}.
They obtain the Cram\'{e}r-Rao type bound, i.e., the lower bound of the trace of the mean square error matrix with an arbitrary weight matrix,
based on the RLD Fisher information matrix.
They showed that the heterodyne measurement $ \{ \frac{|\alpha \rangle \langle \alpha | }{\pi} \} $ achieves the bound and it is optimal.
This measurement is optimal also in an asymptotic sense, which was shown by Hayashi~\cite{Hayashi2000}.

%
%
Here, we consider the prediction problem in the Bayesian framework.
Assume that unknown parameter $\theta $ is distributed subject to
\begin{equation}
\pi(\theta)= \frac{1}{2\pi \tau^2}\exp \left(-\frac{ | \theta - \xi |^2 }{2\tau^2} \right), \label{eq:prior}
\end{equation}
where $\xi \in \complex , \tau^2 > 0 $ are so-called hyperparameter. 

%
%
%
In this section we only consider $n=m=1$ case for simplicity.
General case for arbitrary $n$ and $m$ is considered in the next section.  
When $n=1$, it is natural to adopt the heterodyne measurement above.
Then the estimator of $\theta $ is given by $\hat{\theta}(\alpha) = \alpha$, 
where the measurement outcome $\alpha $ is distributed by
\[
\alpha \sim p^{M} (\alpha | \theta) = \frac{1}{\pi(N+1)} \exp \left(-\frac{|\alpha - \theta|^2}{N+1}\right).
\] 

%
%
%
We calculate the average relative entropy for two predictive density operator $\rho_{\hat{\theta}}$ and ${\rho}_{\pi}$.
Straightforward calculation yields
\begin{eqnarray*}
\rho_{\hat{\theta}(\alpha)} &=& \frac{1}{\pi N}  \int_{\complex} \exp \left( - \frac{ | \beta -\alpha |^2   }{N} \right) | \beta \rangle \langle \beta | 
  \rmd ^2 \beta, \\
{\rho}_{\pi}(\alpha ) &=& \frac{1}{\pi (N+2\Delta^2)} \! \int_{\complex} \!\exp \! \left( - \frac{ | \beta -\bar{\theta} |^2   }{N+2\Delta^2} \right) | \beta \rangle \langle \beta | 
  \rmd ^2 \beta,  
\end{eqnarray*}
where 
\[
\bar{\theta} \!:=\! \frac{ \left( \!\frac{N+1}{2} \right)\!^{-1}\alpha  +\! (\tau^2)\!^{-1} \xi  }{ \left(\frac{N+1}{2}\right)^{-1} \!+\! (\tau^2)^{-1}  }, 
 \ (\Delta^{2})^{-1}\! := \!\left(\!\frac{N+1}{2} \!\right)\!^{-1} \!+ (\tau^{2})\!^{-1}.
\]
The average relative entropy for them is also obtained by
\begin{eqnarray*}
\mathcal{R}_{p}&:=& E^{\pi}E^{M} [D(\rho_{\theta} || \rho_{\hat{\theta}} )] 
 = (N+1)\log \left( \frac{N+1}{N} \right), \\
\mathcal{R}_{\pi}&:=& E^{\pi}E^{M}[D(\rho_{\theta} || \hat{\rho}_{\pi} )] \\
 &=& \log \frac{1}{N+1} + N \log \frac{N}{N+1}  - \log \frac{1}{N\!+ \!2\Delta^2 \!+ \!1}  \\
 && {}- (N \!+ \! 2 \Delta^2) \log \frac{N+\! 2\Delta^2}{ N+ \!2 \Delta^2\! +\!1 },
\end{eqnarray*}
where we used the formula for the Gaussian states family
\begin{eqnarray}
 D(\rho_{\zeta, N} || \rho_{\zeta', M}) 
\!=\! \log\left( \frac{M+1}{N+1} \right) \!+\! N\log \left( \frac{N}{N+1}\frac{M+1}{M} \right) &&  \nonumber \\
	  + \log\left( \frac{M+1}{M} \right) |\zeta - \zeta'|^2. \quad \quad \quad \quad \quad \  && \label{rela}
\end{eqnarray}
%
%
Since $\mathcal{R}_{\pi}$ is monotone increasing with $\tau^2$,
\begin{eqnarray*}
\sup_{\tau^2>0} \mathcal{R}_{\pi} &=& \lim_{\tau^2 \rightarrow \infty} \mathcal{R}_{\pi} \\
 &= &\log \frac{1}{N+1} + N \log \frac{N}{N+1} \\
&&- \log \frac{1}{2N+2} - (2N+1) \log \frac{2N+1}{2N+2} \\
 &\equiv &\mathcal{R}_{*}. 
\end{eqnarray*}
In addition, from the straightforward calculation we can show $\mathcal{R}_{p} >  \mathcal{R}_{*} \geq  \mathcal{R}_{\pi}$.
Thus, it is shown that the Bayesian predictive density operator ${\rho}_{\pi}$ is better than the plug-in density operator $\rho_{\hat{\theta}}$ based on $\hat{\theta}(\alpha)$.

%
%
Since the model $\mathcal{M}$ is translation invariant, it seems natural to adopt 
the Lebesgue measure $\pi_{J}(\theta ) \rmd^2 \theta \propto \rmd^2 \theta $ as a noninformative prior.
Although $\int \pi_J (\theta) \rmd^2 \theta = \infty$, as classical statistics, various quantities are obtained by taking the limit $\tau^2 \rightarrow \infty$.
Since $2 \Delta^2 = N+1$, Bayesian predictive density operator is given by
\[
{\rho}_{\pi_{J}} = \frac{1}{\pi (2N+1)}  \int_{\complex} \exp \left( - \frac{ | \beta -\bar{\theta} |^2   }{2N+1} \right) | \beta \rangle \langle \beta | 
  \rmd ^2 \beta,  
\]
and the average relative entropy is equal to $\mathcal{R}_{*}(< \infty)$.

\section{Prediction of unknown Gaussian state from $n$ sample}

Now, we deal with more general case.
Assume that the unknown $n+m$ systems $\rho_{\theta, N}^{\otimes (n+m)}$ are prepared, where $N$ is known and $\theta $ is unknown 
and subject to the prior (\ref{eq:prior}).
We fix the heterodyne measurement $\{  \left( \frac{|\alpha \rangle \langle \alpha | }{\pi} \right)^{\otimes n} \}$ and 
perform it for arbitrarily chosen $n$ systems.
Then each data $\alpha_i  \in \complex $ is independently subject to 
\[
\alpha_i  \sim p^{M} (\alpha_i  | \theta) = \frac{1}{\pi(N+1)} \exp \left(-\frac{|\alpha_i  - \theta|^2}{N+1}\right).
\] 
We consider the estimation of the remaining $m$ systems, $\sigma_{\theta} := \rho_{\theta}^{\otimes m}$ from these data $(\alpha_1, \dots, \alpha_n)$.
Let us calculate the average risk for $\sigma_{\hat{\theta}} = \rho_{\hat{\theta}}^{\otimes m} $
 and ${\sigma}_{\pi} (\alpha) $.
%
%
%
The plug-in density operator is given by
\begin{eqnarray*}
\sigma_{\hat{\theta}(\alpha)} &=& \rho_{\hat{\theta}(\alpha)}^{\otimes m} \\
&=& \left\{  \frac{1}{\pi N}  \int_{\complex} \exp \left( - \frac{ | \beta -\bar{\alpha}_n |^2   }{N} \right) | \beta \rangle \langle \beta | 
  \rmd ^2 \beta  \right\}^{\otimes m},
\end{eqnarray*}
where 
\[
\bar{\alpha}_n := \frac{1}{n}\sum_{i=1}^{n}\alpha_i
\]
 is a maximum likelihood estimator.
On the other hand the Bayesian predictive density operator is given by
\begin{equation}
 \sigma_{\pi}(\alpha) = \int \rmd^2 \beta_1 \cdots \rmd^2 \beta_m \bigotimes_{j=1}^{m} |\beta_j \rangle \langle \beta_j | p_{\pi}(\beta | \alpha), \label{eq:bpd}  
\end{equation}
where  
\[
p_{\pi}(\beta | \alpha) =  \frac{1}{\pi (N + 2m \Delta_{n}^2 ) } \left( \frac{1}{\pi N}\right)^{m-1}
			\exp \left(- \half \frac{1}{\tilde{\Delta}^2_{n,m}} B \right) 
\]
and 
\[
B := p ( |\beta_1 |^2 + \cdots + |\beta_m |^2 ) + q |\bar{\theta}|^2
 - | p(\beta_1 + \cdots + \beta_m ) + q \bar{\theta} |^2,
\]
\[
 (\tilde{\Delta}_{n,m}^{2})^{-1} := (\Delta^{2}_{n})^{-1} + m \left( \frac{N}{2} \right)^{-1},
\] 
\[
 p := \frac{ (\frac{N}{2})^{-1} }{ (\Delta^{2}_{n})^{-1}  + m (\frac{N}{2})^{-1} }, \ q: 
 =\frac{ (\Delta^{2}_{n})^{-1}  }{ (\Delta^{2}_{n})^{-1}  + m (\frac{N}{2})^{-1} }, 
\]
where 
\begin{eqnarray*}
\bar{\theta}(\alpha ) &:=& \frac{ \left( \!\frac{N+1}{2} \right)\!^{-1} \sum_{i=1}^{n} \alpha_i  +\! (\tau^2)\!^{-1} \xi  }{ n \left(\frac{N+1}{2}\right)^{-1} \!+\! (\tau^2)^{-1}  }, 
  \\
 (\Delta_n^{2})^{-1}\! &:=& n \left(\!\frac{N+1}{2} \!\right)\!^{-1} \!+ (\tau^{2})\!^{-1}.
\end{eqnarray*}
Since
\begin{eqnarray*}
 &&\Exp^{\pi} \Exp^{M} | \bar{\alpha}_n - \theta |^2 =\frac{N+1}{n}\\
 \mbox{and} \ &&\Exp^{\pi} \Exp^{M} | \bar{\theta}(\alpha) - \theta |^2 = 2 \Delta^2_n, 
\end{eqnarray*}
each average relative entropy is obtained by
\begin{eqnarray*}
\mathcal{R}_{p}&:=& E^{\pi}E^{M} [D(\sigma_{\theta} || \sigma_{\hat{\theta}} )] 
 = \frac{m}{n} (N+1)\log \left( \frac{N+1}{N} \right), \\
\mathcal{R}_{\pi}&:=& E^{\pi}E^{M}[D(\sigma_{\theta} || {\sigma}_{\pi} )] \\
 &=& \log \frac{1}{N+1} + N \log \frac{N}{N+1}  \\
 && {} - \log \frac{1}{N+2m\Delta_n^2 + 1} \\
 &&    {} - (N+ 2m \Delta_n^2) \log \frac{N+2m\Delta_n^2}{ N+2 m \Delta_n^2 +1 },
\end{eqnarray*}
Again it is easily shown that $\mathcal{R}_{p} > \mathcal{R}_{\pi}$ for arbitrary hyperparameter $\xi $ and $\tau^2 > 0$.

\section{Concluding remarks}


%
%
Strictly speaking, the proof of theorem 1 is valid only for finite-dimensional cases (i.e., $\dim \Hilbert < \infty$)~\cite{FT3}.
Thus, we only compare the plug-in predictive density operator based on the maximum likelihood estimate $\hat{\theta}$ and
 the Bayesian predictive density operator and show that the latter is better than the former in the average relative entropy.
However, we expect that theorem 1 can be extended to infinite-dimensional cases under some regularity conditions
 such as the exchangeability of the order of $\Tr $ and $\int \dtheta \ \pi(\theta )$ and integrability of $\rho_{\pi}(x) = \int \dtheta \ \pi(\theta | x ) \rho_{\theta }$.  
The quantum Gaussian states family is known to have good properties as the classical Gaussian family has~\cite{Holevo}.
Therefore, it could be shown that the Bayesian predictive density operator is really the best predictive density under the prior (\ref{eq:prior}).
Such rigorous arguments is left for future study.

\appendix

\section{Calculation of the formula (\ref{rela})}

In this section, we derive the formula (\ref{rela}).
First we review the notation and the mathematical description that we need to show the formula (\ref{rela}).
For details of them and physical meaning, see, e.g., Walls and Milburn~\cite{Walls} .

Recall that the projective unitary representation of the translation group on the complex plane is given by 
\[
 U_{\theta} U_{\eta} = e^{\im (\theta \bar{\eta})} U_{\theta +\eta}, \ \forall \theta, \eta \in \complex,
\]
and the Gaussian state with mean parameter $\theta$ is given by unitary transformation of this group,
\[
 \rho_{N, \theta} = U_{\theta} \rho_{N,0} U_{\theta}^{*}, \ \forall \theta \in \complex,
\] 
where $U^{*}$ denotes the adjoint operator of $U$.(It is often denoted as $U^{\dagger}$ in physics.)
On the other hand, coherent state vector is defined in the following form,
\[
  | \alpha \rangle = e^{- \frac{ |\alpha|^2}{2} } \sum_{n=0}^{\infty} \frac{\alpha^n}{ \sqrt{n!} } | n \rangle,
\]
Here, $| n \rangle$ denotes $n$ photon excited state, which is defined  by
\[
 | n \rangle := \frac{ (a^{*})^{n}}{  \sqrt{n} } | 0 \rangle,
\]
where $a^{*}$ is so called creation operator. 
Please note that $ \langle n | m \rangle = \delta_{nm}, n,m=0,1,\dots $ and 
\begin{eqnarray*}
 \langle n | \alpha \rangle &=& \langle n | \left\{  \sum_{n=0}^{\infty} \frac{ \alpha^m}{\sqrt{m!}}| m \rangle \times e^{-\half |\alpha|^2 }  \right\} \\  
	&=&  \frac{ \alpha^n}{\sqrt{n!}}| n \rangle \times e^{-\half |\alpha|^2 }
\end{eqnarray*}
Coherent state vector is a mathematical representation of  the light of a certain frequency.

Now we derive the formula.
The key point is to calculate the following trace. 
\begin{eqnarray*}
\Tr \rho_{N, \theta} \log \rho_{M, \eta} &=&
  \Tr \rho_{N,\theta} \log ( U_{\eta} \rho_{M,0} U_{\eta}^{*} ) \\
 &=&  \Tr \rho_{N,\theta} U_{\eta} ( \log  \rho_{M,0} ) U_{\eta}^{*}  \\
 &=&  \Tr U_{\eta}^{*}  \rho_{N,\theta} U_{\eta} ( \log  \rho_{M,0} ) \\
 &=&   \Tr \rho_{N,\theta -\eta} ( \log  \rho_{M,0} ). \\
\end{eqnarray*}
For simplicity, we calculate  $\Tr \rho_{N,-\eta} ( \log  \rho_{M,0} )$.
Recall that the $\rho_{N,0}$ is diagonalized with the orthonromal basis $\{ | n \rangle \}_{n=0, 1, \dots}$,
\[
 \rho_{N, 0} = \sum_{n=0}^{\infty} \frac{1}{N} \left( \frac{N}{N+1}\right)^{n+1} | n \rangle \langle n | .
\]
We obtain the logarithm of this density operator.
\[
 \log \rho_{M,0} = \sum_{n=0}^{\infty} \log  \left\{ \frac{1}{M} \left( \frac{M}{M+1}\right)^{n+1}  \right\}  | n \rangle \langle n | 
\]
and the matrix element with coherent vector $|\alpha \rangle$ is given by
\begin{eqnarray*}
&&\langle \alpha | \log \rho_{M,0}  |\alpha \rangle \\
&=& \sum_{n=0}^{\infty} \log  \left\{ \frac{1}{M} \left( \frac{M}{M+1}\right)^{n+1}  \right\} 
 | \langle \alpha | n \rangle |^2  \\
 &=&  \sum_{n=0}^{\infty}  \left\{  \log \frac{1}{M+1}  +   n  \log \left( \frac{M}{M+1} \right)   \right\} 
 \frac{  ( |\alpha|^{2}  )^n }{ n! } e^{-|\alpha|^2} \\
 &=& \log  \frac{1}{M+1}  +    |\alpha|^2  \log \left( \frac{M}{M+1} \right).  
\end{eqnarray*}
Then,
\begin{eqnarray*}
 && \Tr \rho_{N,-\eta}\log \rho_{M,0} \\
 &=& \!\!\! \int_{\complex}\frac{1}{\pi N} \exp \left\{ - \frac{1}{N}| \alpha + \eta |^2 \right\}
	\langle \alpha | \log \rho_{M,0}  |\alpha \rangle \\
 &=& \!\!\! \int_{\complex}\frac{1}{\pi N} \exp \left\{ - \frac{1}{N}| \alpha + \eta |^2 \right\} \\  
  && {} \times \left\{   \log  \frac{1}{M+1}  +    |\alpha|^2  \log \left( \!\! \frac{M}{M+1} \right) \!\! \right\}  \\
  &=&   \log  \frac{1}{M+1}  + \log \left( \frac{M}{M+1} \right)  
	\{  |\eta|^2 + N   \}  \\
\end{eqnarray*}

Using this formula, we obtain the relative entropy formula (\ref{rela}).
\begin{eqnarray*}
 &&D( \rho_{N, \theta} || \rho_{M, \eta}) \\
&=& \Tr \{ \rho_{N,\theta} ( \log \rho_{N, \theta} - \log \rho_{M,\eta} ) \} \\
 &=& \Tr \{  \rho_{N,0}  ( \log \rho_{N, 0}) \}  -  \Tr \{ \rho_{N, \theta- \eta} (\log \rho_{M,0} ) \} \\
 &=& \log  \frac{M+1}{N+1}  + N \log \left( \frac{N}{N+1}  \frac{M+1}{M}\right)  \\
	&& {}  -  \log \left( \frac{M}{M+1} \right)  |\theta -\eta|^2.
\end{eqnarray*}

\begin{acknowledgments}
F.T. was supported by JSPS.
\end{acknowledgments}

\newpage 
\bibliographystyle{apsrev}
\bibliography{myonly}

\end{document}